\documentclass[sigconf, authorversion]{acmart}

\usepackage{graphicx}
\usepackage{hyperref}
\usepackage{amsmath}
\usepackage{array}
\usepackage{pdfpages}
\usepackage{color,soul}

\copyrightyear{2025}
\acmYear{2025}
\acmConference[Preprint]{}{February 2025}{}
\acmBooktitle{Preprint} 
\acmDOI{10.1145/3706598.3714014}
\acmISBN{979-8-4007-1394-1/25/04}

\begin{document}

\title[Scaffolding Empathy: Training Counselors with Simulated Patients and Performance Visualizations]{Scaffolding Empathy: Training Counselors with Simulated Patients and Utterance-level Performance Visualizations}

\author{Ian Steenstra}
\affiliation{%
  \institution{Northeastern University}
  \city{Boston, MA}
  \country{USA}}
\email{steenstra.i@northeastern.edu}

\author{Farnaz Nouraei}
\affiliation{%
  \institution{Northeastern University}
  \city{Boston, MA}
  \country{USA}}
\email{nouraei.f@northeastern.edu}

\author{Timothy W. Bickmore}
\affiliation{%
  \institution{Northeastern University}
  \city{Boston, MA}
  \country{USA}}
\email{t.bickmore@northeastern.edu}

\renewcommand{\shortauthors}{Steenstra, et al.}

\begin{abstract}
Learning therapeutic counseling involves significant role-play experience with mock patients, with current manual training methods providing only intermittent granular feedback. We seek to accelerate and optimize counselor training by providing frequent, detailed feedback to trainees as they interact with a simulated patient. Our first application domain involves training motivational interviewing skills for counselors. Motivational interviewing is a collaborative counseling style in which patients are guided to talk about changing their behavior, with empathetic counseling an essential ingredient. We developed and evaluated an LLM-powered training system that features a simulated patient and visualizations of turn-by-turn performance feedback tailored to the needs of counselors learning motivational interviewing. We conducted an evaluation study with professional and student counselors, demonstrating high usability and satisfaction with the system. We present design implications for the development of automated systems that train users in counseling skills and their generalizability to other types of social skills training. 
\end{abstract}



\begin{teaserfigure}
\centering
\includegraphics[scale=0.4]{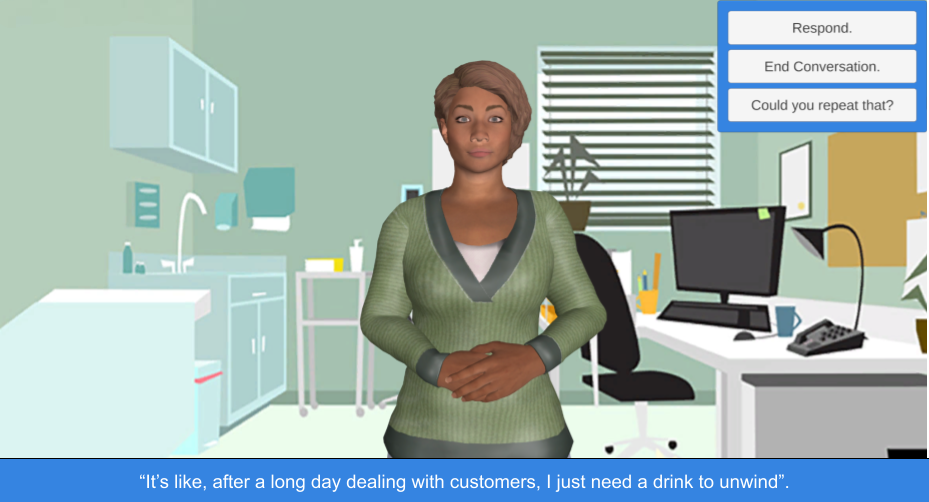}
  \caption{SimPatient: Interaction Interface}
  \Description{This figure shows the interface where counselors interact with the SimPatient. It includes a simulated patient visualized as an animated character and buttons for replying or ending the conversation.}
  \label{fig:interaction}
\end{teaserfigure}

\maketitle

\section{Introduction}
Substance misuse is a major source of morbidity and mortality worldwide, with over 2.5 million deaths attributable to alcohol consumption alone in 2019 \cite{WHO_Alcohol_2024}. Therapeutic counseling is one of the most effect treatments for substance misuse, with motivational interviewing (MI) one of the most widely used counseling frameworks \cite{miller2023motivational}. Motivational interviewing is a collaborative, client-centered approach to enhance motivation for the change of health behavior, through the elicitation and reinforcement of the client's change talk \cite{moyers2016motivational}. Good counselors must acquire the ability to apply theoretical knowledge, understand interpersonal dynamics, and appropriately apply specific counseling skills in real-time during nuanced conversations with clients. Counselors must also develop deep empathy, active listening skills, and the ability to maintain emotional boundaries.
 
Learning these counseling skills is difficult and is primarily taught using traditional classroom methods and media. Maintaining competency, however, requires ongoing professional development, including refresher training for experienced counselors to meet continuing education requirements, often expressed in continuing education units mandated by licensing boards and professional organizations \cite{shiri2023role, USDHHS2024CEU}. One meta-review of 28 studies of counselor training programs for motivational interviewing found that most programs used didactic instruction, half of the programs used role plays, and that many of the programs (18\%) used "standardized patients" (actors) for trainees to practice with, with feedback provided by the instructor or peer trainees \cite{madson2009training}. Although essential for learning, feedback from role-plays and standardized patients is infrequent, inconvenient, and potentially costly, since they involve another person who is often an actor with specialized training, and laborious coding and scoring of trainee-client interactions following standardized protocols, such as those laid out by the Motivational Interviewing Treatment Integrity (MITI) coding manual \cite{moyers2016motivational}.

To improve the effectiveness and efficiency of training counselors, human standardized patients could be replaced with virtual, simulated patients, that are always available for as many practice sessions as needed. While client simulations have been developed for this purpose - both in research \cite{wang2024patient, yosef2024assessing} and commercial products \cite{cutts2020enhancing, grenvik2004resusci} - these have narrowly scripted scenarios and highly constrained actions that trainees can take, limiting their effectiveness. Recent advances in large language models (LLMs) now afford the potential to develop patient simulations that trainees can interact with in an unconstrained, unprompted manner, more closely approximating the experience of a human-human role-play. In addition, LLMs can be used to provide automated performance feedback - both through standard counseling observational metrics and more detailed information about the state of the simulated patient, updated after every trainee utterance.
 
Our goal in this work is to explore the development and evaluation of such a simulated patient training system, which we call SimPatient. Recognizing the need for effective training across all experience levels, this system is designed to benefit both novice trainees, such as counseling students, and experienced professionals who may benefit from continuing education. Following other recent work on the use of LLMs for patient simulation \cite{wang2024patient}, we use an LLM in conjunction with a patient cognitive model to drive our simulation. We also use LLMs to "observe" a trainee interacting with a simulated patient in real-time and provide expert feedback using standard observational assessment protocols and data visualization techniques. Our cognitive model focuses on four key cognitive factors, informed by substance-misuse research \cite{copersino2017cognitive}, deemed crucial for understanding and responding to individuals struggling with alcohol misuse:
\begin{itemize}
\item \textbf{Control:} Represents the patient's perceived ability to regulate their thoughts, emotions, and actions, particularly related to alcohol cravings and consumption.
\item \textbf{Self-Efficacy:} Reflects the patient's confidence in resisting cravings, coping with triggers, and achieving recovery goals.
\item \textbf{Awareness:} Represents the patient's insight into their thoughts, feelings, and behaviors related to alcohol use, including recognizing problematic patterns and potential consequences.
\item \textbf{Reward:} Reflects the degree to which alcohol and its associated cues trigger cravings and automatic behaviors in the patient, highlighting the reinforcing properties of the substance.
\end{itemize}
 
In the rest of this paper, we first describe a requirements elicitation study in which we interview stakeholders about features that would be important to them in such a training system. We then present our resulting design, followed by an evaluation study with stakeholders. Our test domain is motivational interviewing for alcohol misuse.

This research addresses the following research questions:
\begin{itemize}
\item \textbf{R1:} What design features do students and MI experts desire in an MI training and skill assessment system?
\item \textbf{R2:} How do students and MI experts perceive and respond to using a MI training and skill assessment system?
\item \textbf{R3:} To what extent can LLMs effectively simulate patient personas encompassing sociodemographics and personality type?
\item \textbf{R4:} To what extent can LLMs effectively simulate dynamic patient cognitive models, including factors such as self-efficacy?
\end{itemize}

\section{Related Work}

\subsection{Intelligent Systems for Clinical Skill Training}

The field of human-computer interaction has recently seen a myriad of intelligent systems that assist clinicians and healthcare providers in improving their skills and enhancing clinical task execution. 

For example, the KiTT tablet-based learning system was designed to help nursing care students learn kinesthetic-based patient transfers. This system provides instructions and feedback, improving the learning experience in a training room context and showing potential in preserving nurses' health by promoting ergonomic patient transfers \cite{durr2021kitt}. In an attempt to maximize knowledge transfer from training to the real world, Zhou et al. created a 3D projection-based augmented reality system for nursing students, evaluated in a stroke simulation scenario, shown to enhance the realism of patient mannequins \cite{zhou2022bringing}. Similarly, Zadow et al. developed an interactive multitouch tabletop system for medical education for real-time diagnosis and treatment simulations, offering cost-effective and engaging learning environments compared to full-scale mannequins \cite{von2013simmed}. These studies underscore the benefits of realistic educational simulations and the trade-off between realism and cost.

The most relevant to our system is ConverSense, which uses machine learning to process social signals from patient-provider interactions to provide feedback to clinicians, facilitating self-reflection and communication improvement. Healthcare providers indicated the potential for long-term communication benefits from using ConverSense \cite{bedmutha2024conversense}.

\subsection{Training Systems for Counseling Conversations} 
The system proposed by Albright et al. \cite{albright2016harnessing} uses MI techniques to guide users through role-playing scenarios with embodied agents, focused on initiating conversations related to mental health. This system prepares individuals to engage in meaningful conversations and provides them with feedback based on their performance.
Similarly, Murali et al. created a system that involves role-playing with a conversational agent to teach counseling skills to laypersons in the context of vaccination promotion \cite{muraliTrainingLayCounselors2022b}. Several commercial products have also been developed that use conversational agents for the training of health conversation skills. These have been deployed and evaluated in various application areas, including prevention of suicide among college students \cite{rein2018}, training for healthcare providers and patients to talk about overuse of antibiotics \cite{schoenthaler2017}, and training for nurses to conduct screening for adolescent substance use \cite{burmester2019}.

\subsection{Standardized Patients and Patient Simulation Systems}
Traditionally, standardized patients--actors who allow providers to practice clinical skills by enacting real-life scenarios as a patient--have been widely used in healthcare to enhance medical education, primarily for communication and clinical skills. Prior research indicates that the use of standardized patients allows trainees to practice real-life scenarios in a controlled environment, fostering a sense of realism and emotional engagement that traditional educational methods may lack. Previous work has emphasized the effectiveness of standardized patients in medical training by showing that engaging with standardized patients leads to better acquisition of communication skills compared to didactic methods \cite{lane2007use}. However, using standardized patients can be costly and time consuming, due to cumbersome recruiting, training, and standardization of their interactions \cite{flanagan2023standardized, whitaker2015motivational, swanson2013assessment}.

Emerging technologies, including chatbots and embodied conversational agents, have significantly transformed the patient simulation process. The introduction of simulated patients provides flexibility in training by allowing learners to interact with computerized patients integrating natural language processing and realistic emotional responses \cite{hubal2000virtual}. These technologies not only reduce costs associated with hiring actors, but also enhance the availability and adaptability of training scenarios, allowing personalized and context-rich interactions for clinicians. One such approach includes patient-focused simulation, which integrates real human patients with high-fidelity simulators, fostering essential communication and decision-making skills alongside technical proficiency \cite{bartlett2021high}. Studies indicate that more realistic simulations lead to improved retention of skills and knowledge, enhanced engagement, and greater confidence in clinical settings \cite{kneebone2006human}.  

Overall, the move towards the integration of human-like interaction has been shown to be crucial to bridge the gap between theoretical education and practical application in healthcare  \cite{bartlett2021high, kneebone2006human, sarker2007simulation}

\subsection{LLMs for Simulated Patient Dialogue}
Recent advancements in LLMs have made them an attractive candidate to drive simulated patient dialogue and system reasoning for training purposes. To this end, CureFun utilizes multiple LLMs to simulate patients for medical diagnosis education, and offers automated feedback on trainee performance by analyzing medical scenarios to inform its responses \cite{li2024leveraging}. Similarly, \citeauthor{liao2024automatic} introduce State-Aware Patient Simulator to bridge the gap between static medical knowledge assessments and dynamic clinical interactions \cite{liao2024automatic}. Their approach allows for more realistic LLM evaluation in multi-turn doctor-patient simulations.

Despite impressive conversational abilities of LLMs, ensuring accurate and nuanced portrayals of specific patient populations requires expert input. \citeauthor{louie2024roleplay} address this challenge with Roleplay-doh, a pipeline that enables domain experts to guide LLM simulations through elicited principles. This work emphasizes the importance of expert feedback, particularly for sensitive domains such as mental health, and provides a mechanism for incorporating such feedback directly into LLM prompting \cite{louie2024roleplay}. \citeauthor{chen2023llm} explore the use of ChatGPT for the simulation of psychiatrists and patients, developing a dialogue system informed by psychiatrist input \cite{chen2023llm}. Their evaluation with real clinicians highlights the feasibility of LLM-powered chatbots in psychiatric scenarios while emphasizing the importance of careful prompt design for realistic and ethical interactions.

The use of LLMs for social skill training, including conflict resolution and communication, presents a promising application \cite{shaikh2024rehearsal, yang2024social}. For example, \citeauthor{yang2024social} discuss a framework leveraging LLMs to create accessible and engaging social skills training environments \cite{yang2024social}. Their AI Partner, AI Mentor framework combines experiential learning with realistic practice and tailored feedback, which aligns with SimPatient's goal of providing a safe space for practicing MI techniques. However, LLMs, especially those trained with reinforcement learning from human feedback as analyzed by Perez et al. \cite{perez2022discovering}, exhibit a positivity bias, tending towards agreeableness and a desire to comply with user requests. This raises challenges for simulating resistant or deceptive patients, scenarios crucial for robust training \cite{lee2024llms, petrov2024limited, mieleszczenko2024dark}. Further research is needed to overcome these limitations and accurately model the full spectrum of patient behaviors.

Closest to our work is that of Wang et al., who proposed a simulated patient framework to enhance Cognitive Behavioral Therapy training \cite{wang2024patient}. In their work, patient cognitive models, such as emotional states and maladaptive cognitions grounded in the principles of Cognitive Behavioral Therapy, are integrated with LLMs to ensure high-fidelity simulated patient interactions. Our approach differs in that: 
(1) SimPatient uses cognitive models grounded in the literature on substance misuse \cite{copersino2017cognitive}; 
(2) SimPatient allows for multiple complete and natural conversations with a simulated patient, followed by an opportunity for reflection through a comprehensive dashboard; and 
(3) the evaluation dashboard visualizes tracked changes in the simulated patient's cognitive model, allowing users to reflect on what they said and the resultant change in the simulated patient's cognitive model.

\subsection{LLMs for Motivational Interviewing Evaluation}
In recent years, language models, and particularly LLMs, have gained traction in evaluating the quality of counseling conversations from both technical and relational perspectives \cite{li2024automatic, ahmed2022automatic, perez2019makes, imel2019design}. These evaluation modules can be integrated with patient simulation platforms, offering a novel approach to evaluating the performance of counselors in MI sessions. For instance, Yosef et al. introduced an LLM-based digital patient platform that generates diverse patient profiles and utilizes LLMs to rate the quality of therapy sessions based on established frameworks such as the Motivational Interviewing Skills Code and the MITI \cite{yosef2024assessing}. These LLMs were found to effectively distinguish between varying levels of therapist expertise by completing questionnaires designed for human patients, indicating the potential for automated feedback in therapeutic settings.

Researchers have also developed datasets of counseling sessions annotated for both the therapist and client behaviors. Among these, the MI-TAGS dataset allows training language models to perform tasks such as utterance-level annotations and overall session scoring, achieving performance comparable to human annotators \cite{cohen2024motivational}. Using similar datasets, LLMs have been trained as evaluators for therapist responses, providing a way to gauge their adherence to MI principles in real time \cite{wu2023natural, wu2022anno}. In our proposed system, we used a mix of zero-shot and few-shot approaches to utilizing LLMs for evaluating counselor actions in terms of adherence to MI principles, which eliminates the need for data-intensive and time-consuming fine-tuning.

\subsection{LLM-Based Multi-Agent Systems}
The rise of LLMs has fueled the development of multi-agent systems where each agent is driven by an LLM \cite{guo2024large}. These LLM-based multi-agent systems leverage the inherent strengths of LLMs, such as natural language understanding/generation and knowledge representation \cite{pezeshkpour2023measuring}, to create agents capable of sophisticated interaction and collaboration. Hippocratic AI's Polaris system \cite{mukherjee2024polaris} provides a compelling example of a multi-agent LLM architecture specifically designed for healthcare applications using embodied agents. Polaris's design incorporates specialized support agents to enhance safety and address nuanced medical queries during patient interactions. 

Despite the rapid progress, several research challenges remain in the field of LLM-based multi-agent systems. Effective communication and coordination between agents are important for successful collaboration, requiring the development of specialized communication protocols and coordination mechanisms \cite{liu2023dynamic, zhang2023proagent}. The tendency of LLMs to generate factually incorrect or nonsensical outputs (hallucinations) poses a significant challenge, necessitating techniques for grounding these systems in real-world knowledge or simulated environments \cite{huang2023survey}. Furthermore, running multiple LLMs simultaneously can be computationally expensive, driving the need for research on optimizing LLM-based multi-agent systems for scalability and efficiency \cite{chen2023agentverse}.

\section{Formative Study}
Before developing SimPatient, we conducted a formative study to (1) identify desired design features for an MI training and skill assessment system for both novice trainees and experienced professionals seeking continued education or refresher training and (2) validate the relevance of four cognitive factors (Control, Self-Efficacy, Awareness, and Reward) for simulating patient internal states. We conducted semi-structured interviews with professional counselors experienced in MI and university students in health counseling-related programs. Since we planned to incorporate performance metrics and visualizations, the study aimed to understand stakeholder preferences for specific measures, visualization formats (e.g., graphs, transcripts), and feedback timing (e.g., during or after sessions). We also explored stakeholder perspectives on the chosen cognitive factors, their potential for capturing diverse patient types, and how visualizing their changes could support MI skill development. For example, regarding performance measures, we asked: "The system will provide evaluation measures of how well a user did based on their use of MI skills. What type of measures do you believe to be important to provide to users?" Follow-up questions probed the rationale for chosen measures, preferred visualization styles, and appropriate scoring mechanisms. We then presented participants with two preliminary visualizations: (1) a radial chart depicting "Empathy," "Cultivating Change Talk," and "Softening Sustain Talk" and (2) a bar chart showing the frequency of MI behaviors. This allowed participants to react to concrete examples and suggest improvements or additional visualizations and measures.

\subsection{Study Procedure}
We conducted 30-minute semi-structured interviews with participants. We recruited participants from \textit{Upwork.com}, a platform for freelance professionals, 
and from our university's communication portal to recruit counseling students. Our inclusion criteria for professional counselors were: (1) being at least 18 years old and fluent in English, and (2) having used motivational interviewing in a professional capacity before. Our inclusion criteria for student counselors were: (1) being at least 18 years old and fluent in English, and (2) currently a college student in a psychology, nursing, social work, public health or related program. The IRB of our institution approved the study and participants were compensated for their time.  

The interviews were transcribed verbatim using a professional transcription service and analyzed using thematic analysis \cite{braun2012thematic}.

\subsection{Results}
\subsubsection{Participants}
Our study included 11 participants: 6 professional counselors (4 female, 2 male) and 5 student counselors (4 female, 1 male). The professional counselors ranged in age from 27 to 46 (mean = 35.5, SD = 6.02), and most (5 out of 6) held advanced degrees. Their occupations included school social worker, dietitian, and psychologist. The student counselors ranged in age from 23 to 45 years (mean = 31, SD = 8.07). They were pursuing degrees in Bachelor of Science in Psychology, Master of Science in Counseling Psychology, Ph.D. in Counseling Psychology, and Bachelor of Science in Social Work. All participants except for one college student were familiar with MI and had used it before.

\subsubsection{Qualitative Results}
Our thematic analysis of the design interview transcripts revealed several key design features for our training system. 

\textbf{MI Skills Assessment:} Participants strongly favored evaluation measures rooted in established MI principles and techniques. For instance, empathy and partnership were suggested as metrics: "\textit{partnership, I think that's definitely important...sometimes folks may engage in behaviors, but they may have shame and confronting their issue, and so the ability for the other side to have empathy can...create a safe space for folks to try to seek help} [P2, male professional counselor]", as well as affirmations and reflections: "\textit{considering affirmation, so maybe if there's an ability to track maybe what is more positive rather than negative...I think in regards to motivational interviewing, also probably reflective listening} [P5, female student counselor]". Additionally, participants suggested metrics that were closely aligned with empathy and cultivating change talk, such as "\textit{listening, active listening skills} [P1, female professional counselor]" and "\textit{following the person who's talking into areas where they are willing to make change} [P1, female professional counselor]". 

\textbf{Multi-Modal Evaluation Visualizations:} Participants consistently expressed a desire for a multifaceted evaluation experience, advocating for a mixture of graphical, numerical, and transcript-based visualizations. One professional counselor, emphasizing the value of a holistic view, remarked, "\textit{I am definitely more of a visual learner, so I think personally graphic would be good, but I think having the transcript and being able to really hone in on what you said and how you said it would be fantastic. So maybe a combination of the two} [P11, female professional counselor]". Other participants valued graphs as well, \textit{"I think other things...like graphs and visualizations} [P8, male student counselor]", but also advocated for numerical values (such as percentages and 1-10 scales) as well as the specific MI category that each utterance belongs to in the transcript: \textit{"Alongside the transcript so that they can actually point out where is the change talk? Where is the sustained talk?} [P10, female professional counselor]".  

\textbf{Transparency \& Reasoning for Scores:} A recurring theme throughout the interviews was the desire for transparency and individualized explanations for the assigned evaluation scores. Participants wanted to understand not only their scores but also the reasoning behind them. One student counselor, highlighting the potential for scores to evoke negative reactions, stated: "\textit{unfortunately sometimes people can still take offense to certain numbers... so just having the ability for explanation and leaving things on a lighter note...helps kind of negate a lot of the negativity} [P5, female student counselor]". Additionally, a professional counselor highlighted the desire to understand the reasoning for why an utterance was coded as a certain MI behavior code: \textit{"being able to understand why that specific piece is complex reflection...especially if there's something you're struggling with, a concept that you don't fully grasp yet, and you're like, okay, I see that I did this here, but I don't really understand why this is this thing. You can go back and understand the reasoning.} [P11, female professional counselor]".

\textbf{End-of-Session Evaluation Dashboard:} Although some participants acknowledged the potential value of real-time feedback, particularly in longer training sessions, the majority favored receiving a comprehensive evaluation summary after the session concludes: "\textit{I feel like it would be better to show them at the end of each session and then be able to track their progress throughout the later two to see if you're able to show them feedback...after each one, are they able to improve on those points where they weren't scoring as well} [P7, female professional counselor]". This preference for post-session feedback also reflects a desire to avoid distractions and performance anxiety during the simulated interaction. As one professional counselor articulated, "\textit{I would think after, just because the person in question may respond based on those attributes, and so may change their frame of response.} [P2, male professional counselor]" This sentiment highlights the potential for real-time feedback to disrupt the natural flow of the conversation and introduce biases into user behavior.

\textbf{MI Proficiency Threshold Comparisons:} In addition to personalized feedback, participants expressed a desire to benchmark their performance against established standards of MI proficiency. One professional counselor, emphasizing the value of visualizing their performance relative to these benchmarks, suggested, "\textit{I would have an additional graph that shows the ratios and then just graphs of what each proficiency... looks like in comparison to what they did} [P10, female professional counselor]".

\textbf{Realism Factors:} In order to facilitate a realistic counseling session with a simulated patient, participants advocated for a few factors that would strengthen the realism of the training experience. For instance, one professional counselor suggested the inclusion of a between-session event, "\textit{between each session...what if the AI decides to go out and go on a bender with their friends...and then they come back and they have to talk to you or they're actually eliciting some form of change. I think that would also make it more humanistic and realistic...and could give the practitioner almost like a curve ball type situation} [P7, female professional counselor]". Participants also recognized the importance of nonverbal behaviors for more realistic communication and advocated for its integration into the simulated patient: "\textit{integrating maybe facial features} [P8, male student counselor]", emphasizing the need for a more embodied and human-like experience.

\textbf{Dynamic Cognitive Factors:} 
We asked feedback from participants regarding our four cognitive factors—Control, Self-Efficacy, Awareness, and Reward—that were derived from the substance-misuse literature. The majority of participants supported the integration of these specific cognitive factors, asserting their potential to significantly enhance the realism of the simulated patient and offer critical training insights. One professional counselor affirmed the appropriateness of our selection: "\textit{That's pretty solid. And I think as a starting point, just four, it's perfect... it's broad enough to where it encompasses a lot of clients and it's narrow enough to where it can differentiate between them} [P4, male professional counselor]". When asked about the value of visualizing these changing characteristics, one student counselor acknowledged: "\textit{I think it's helpful for the simulation} [P6, female student counselor]", while a professional counselor highlighted the pedagogical benefit "\textit{if they knew what they're looking for as far as what are some signs that someone's being more self-aware throughout the duration of the session} [P4, male professional counselor]". This ability to witness the impact of their actions on the simulated patient's internal state, often obscured in real-world interactions, was broadly viewed as beneficial for learning and promoting reflection on one's MI ability.

\subsection{Design Outcomes}
The SimPatient system's design is a direct result of the formative study, incorporating feedback from both professional and student counselors. Our overarching goal was to create a system that is both user-friendly and pedagogically sound for trainees of all experience levels. Our design was also informed by prior research exploring the use of chat interfaces for delivering personalized feedback, including work on MI in chatbot contexts \cite{samrose2022mia}. A key finding from our formative study was the strong emphasis participants placed on receiving clear, individualized explanations for evaluation scores and specific guidance on how to improve. This feedback significantly influenced the design of SimPatient's multi-agent architecture (\autoref{sec:agent_architecture}), specifically the incorporation of chain-of-thought prompting \cite{wei2022chain} within the agents to generate justifications for scores and tailored recommendations, promoting transparency and deeper understanding of the evaluation process. This approach aligns with broader research highlighting the importance of personalized feedback in online learning environments \cite{cavalcanti2021automatic, maier2022personalized}.

\textbf{Professional Counselor Feedback:} Professional counselors stressed the need for clinically meaningful metrics, initially citing "active listening," "empathy," and "partnership" as examples. After reviewing the prototype radar chart showcasing the four global MI scores (Empathy, Partnership, Cultivating Change Talk, Softening Sustain Talk), they strongly endorsed these metrics but emphasized the importance of providing clear definitions for each, a suggestion incorporated into the final design of the global MI scoring module by placing definitions above the radar chart. They also advocated for benchmarking features, subsequently included in the final dashboard to enable comparisons against established proficiency standards and contextualize performance within clinical benchmarks, and for a pie chart depicting overall MI adherence, which was also added. Based on this feedback, the original set of cognitive factors was retained, and a dedicated LLM agent was created within the SimPatient system to manage their dynamic changes. Finally, the suggestion for incorporating between-session events, as a means of enhancing realism and introducing unexpected challenges ("curve balls") for deeper learning, led to the inclusion of a dedicated LLM agent responsible for generating plausible events that inform the initial dialogue of subsequent sessions.

\textbf{Student Counselor Feedback:} Student counselors also valued comprehensive metrics but emphasized the importance of clear, accessible feedback, particularly regarding the rationale behind evaluation scores and how to improve. This feedback significantly influenced the design of SimPatient's multi-agent architecture (\autoref{sec:agent_architecture}), specifically the incorporation of chain-of-thought prompting \cite{wei2022chain} within the agents. This enabled the agents to generate justifications for scores and code assignments, promoting transparency and deeper understanding of the evaluation process. For example, a Global Scoring Agent was created to generate both the scores for the radar chart (showcasing Empathy, Partnership, Cultivating Change Talk, and Softening Sustain Talk) and detailed explanations for each score using chain-of-thought reasoning, displayed alongside the visualization. Similarly, in response to students' desire for explanations of MI behavior code assignments, a dedicated Behavior Coding Agent was implemented. This agent generates MI behavior codes for each user utterance and provides accompanying explanations for its coding decisions, all displayed within the annotated transcript on the dashboard. Finally, in response to students' expressed need for guidance on improvement, a dedicated Session Summary Agent was created. This agent leverages the full conversation history and scores generated by other agents to provide a paragraph-long summary highlighting strengths, areas needing improvement, and recommendations for future practice.

Based on the identified design outcomes implemented in SimPatient, the following sections detail each module of the evaluation dashboard and describe the underlying LLM-powered multi-agent system responsible for both driving the simulated patient's verbal and nonverbal communication and generating the data presented in the evaluation dashboard modules.

\begin{figure*}[t]
\centering
\includegraphics[width=\textwidth]{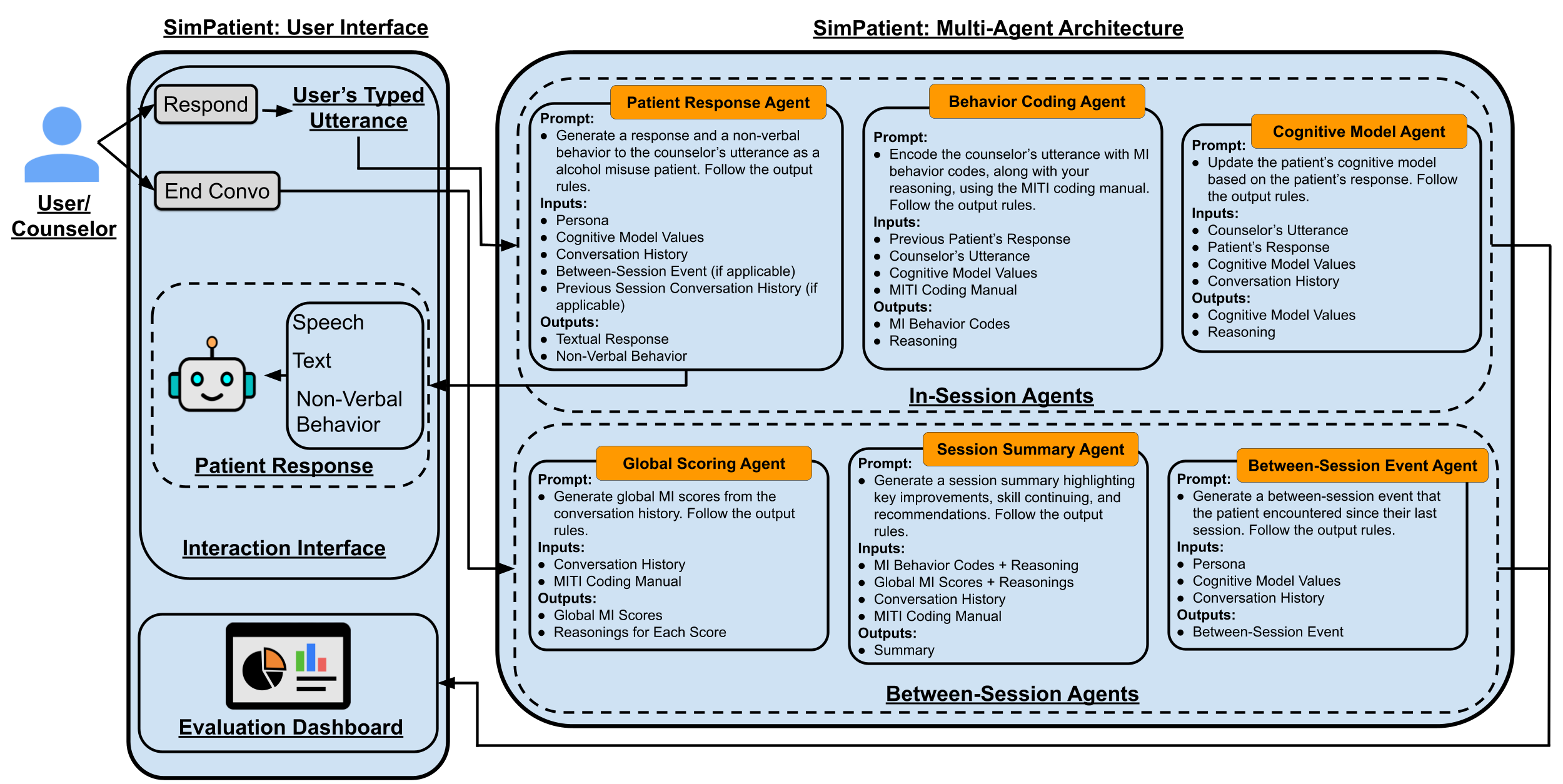} 
\caption{SimPatient Design}
\label{fig:sim_patient}
\Description{This figure provides a high-level overview of SimPatient's architecture. It illustrates the different components of the system, including the in-session and between-session agents, and how they interact with each other and with the user.}
\end{figure*}

\section{SimPatient Design}
Guided by results from our formative study, we translated stakeholder preferences into a concrete system design. Our resulting system, SimPatient (\autoref{fig:sim_patient}), aims to provide a realistic counseling interaction with a simulated patient (\autoref{fig:interaction}), complemented by a comprehensive and tailored evaluation dashboard.

\begin{figure}[htb]
\centering
\includegraphics[width=0.8\columnwidth]{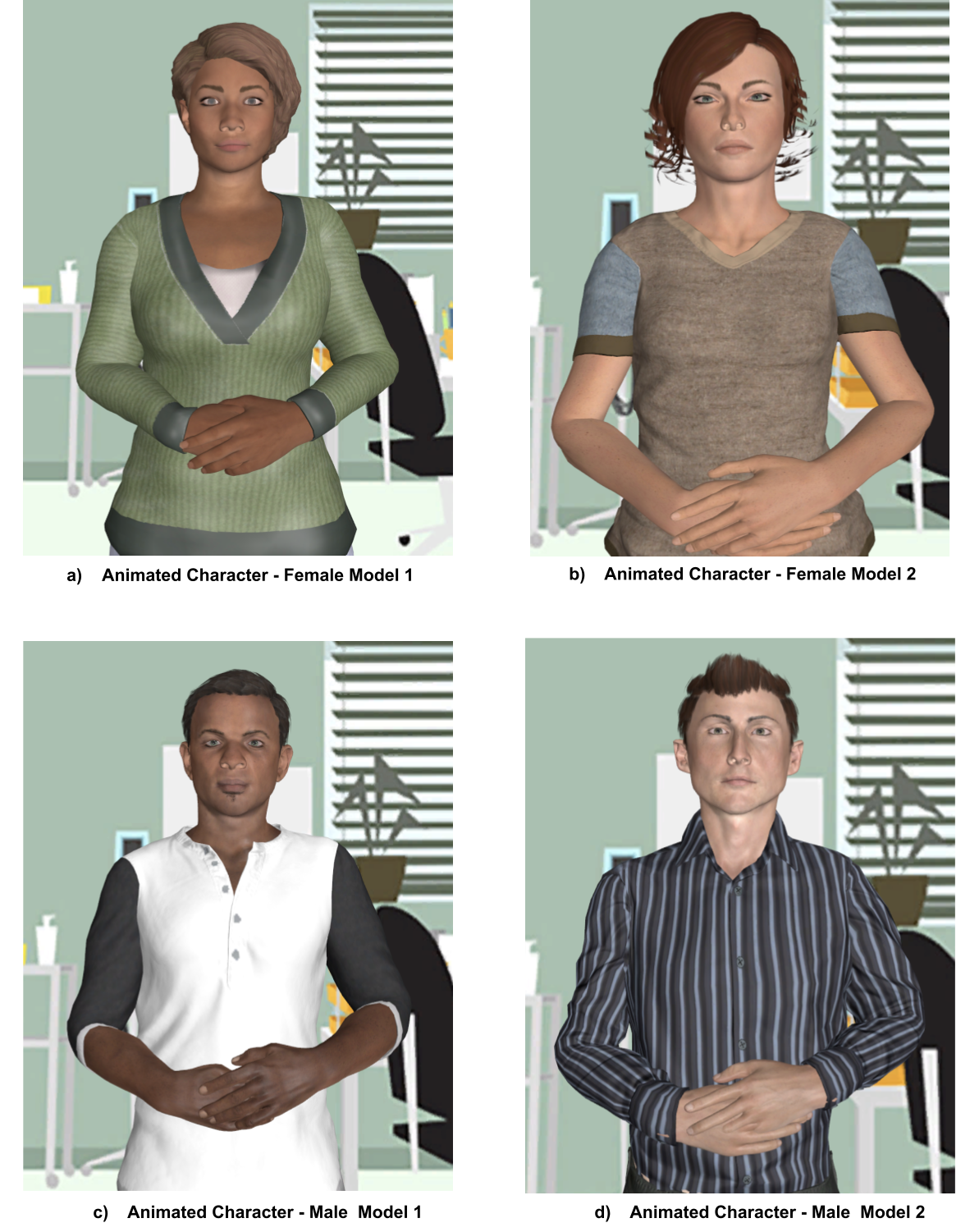}
\caption{Animated Character Model Variations}
\label{fig:animated_characters}
\Description{This figure shows the four different animated character models that counselors can receive to represent their simulated patient.}
\end{figure} 

\subsection{Multi-Agent Architecture}
\label{sec:agent_architecture}
SimPatient employs a multi-agent architecture, similar to systems like Polaris \cite{mukherjee2024polaris}, to power its interactive simulation and evaluation capabilities. This architecture drives both the real-time interaction of the embodied conversational agent, which takes typed text input and produces text-to-speech synthesized verbal responses and animated nonverbal behaviors, and the generation of data displayed in the evaluation dashboard. Each agent within this architecture is realized as a distinct instance of an LLM, receiving specialized prompts tailored to address specific tasks. We use GPT-4o with a temperature setting of 1 for all LLM agents.
\autoref{fig:sim_patient} illustrates the interplay of these agents within SimPatient and our underlying code has been open-sourced \footnote{https://github.com/IanSteenstra/SimPatient}. During the active counseling session, three key agents (in-session agents) operate in concert:
\begin{itemize}
\item \textbf{Patient Response Agent:} This agent is the core of our SimPatient system, generating realistic and contextually appropriate verbal and nonverbal patient responses. It leverages several key inputs: a comprehensive patient persona profile, the current session history, the user's most recent utterance, and, when applicable, the previous session's history and a between-session event summary. 
\item \textbf{Behavior Coding Agent:} This agent analyzes each user utterance, assigns relevant MI behavior codes from the MITI coding manual \cite{moyers2016motivational}, and provides justifications for its coding decisions using chain-of-thought prompting. This fine-grained analysis facilitates utterance-level feedback on the user's MI technique, presented both through the frequency visualization of MI behaviors (Figure \ref{fig:eval_dashboard_example}b) and within the session transcript, where MI behavior codes and their corresponding rationales are displayed alongside each user utterance.
\item \textbf{Cognitive Model Agent:} To simulate the dynamic nature of patient internal states, this agent updates the simulated patient's cognitive factors (Control, Self-Efficacy, Awareness, and Reward),  after every user utterance. By considering both the user's utterance and the simulated patient's response, the agent generates nuanced shifts in these factors and its reasoning for them using chain-of-thought prompting, providing insight into the impact of specific MI techniques which can be viewed through a line graph (\autoref{fig:cognitive_factors_graph}), as well as alongside each patient utterance in the transcript.
The agent assigns a score (1-10) to each cognitive factor, facilitated by its prompt providing explanations of each cognitive factor and example patient utterances representing scores of 1 and 10 (e.g., for self-efficacy: explanation—"Your level of confidence in your ability to resist cravings, cope with triggers, and achieve your recovery goals", 1—"I don't think I can do this. Alcohol has such a hold on me, I always go back to it", 10—"I'm confident I can handle any situation without needing alcohol. I've got this").
\end{itemize}
Upon session conclusion, three additional agents (between-session agents) come into play:
\begin{itemize}
\item \textbf{Global Scoring Agent:} This agent generates global MI scores on a scale of 1 to 5 for "Partnership", "Empathy", "Cultivating Change Talk", and "Softening Sustain Talk", by considering the complete session transcript and established benchmarks from the attached MITI coding manual \cite{moyers2016motivational}. Along with each global MI score, the agent simultaneously generates a rationale using chain-of-thought prompting. These rationales are displayed below the radar chart visualization of the scores (Figure \ref{fig:eval_dashboard_example}a).
\item \textbf{Session Summary Agent:} Using global MI scores, conversation history, and all other calculated MI measures, this agent produces a concise and insightful summary of user performance, highlighting strengths, areas for improvement, and actionable recommendations for future practice.
\item \textbf{Between-Session Event Agent:} To enhance realism and simulate the passage of time, this agent generates a plausible event related to the patient's recovery journey that occurs between sessions, for example a relapse episode at a party.
\end{itemize}

All agents, except the \textbf{Between-Session Event Agent}, deliver their output to the evaluation dashboard, providing users with a comprehensive and multifaceted assessment of their MI skills.

\begin{table*}[t]
\centering
\caption{Inter-rater Reliability and Average Scores for Individual Agents}
\label{tab:agent_validation}
\begin{tabular}{lccc}
\toprule
Agent & Mean Score & Standard Deviation & ICC (95\% CI) \\
\midrule
Patient Response & 4.63 & 0.59 & 0.78 (0.53-0.91) \\
Behavior Coding & 4.70 & 0.46 & 0.77 (0.50-0.90) \\
Cognitive Model & 4.63 & 0.70 & 0.85 (0.67-0.94) \\
Global Scoring & 4.30 & 0.91 & 0.82 (0.60-0.93) \\
Session Summary & 4.70 & 0.82 & 0.93 (0.83-0.97) \\
Between-Session Event & 4.83 & 0.38 & 0.83 (0.63-0.93) \\
\bottomrule
\end{tabular}
\end{table*}

\subsubsection{Preliminary Validation of System Components}
To provide a preliminary validation of each agent within our multi-agent architecture, two authors evaluated 20 hand-coded examples for each agent and rated its output on a 1 to 5 numerical rating scale assessing the overall quality of the agent's output, where 1=Unacceptable (output is completely inaccurate, implausible, and/or nonsensical) and 5=Excellent (output is highly accurate, perfectly plausible, and of excellent quality). A two-way mixed effects, absolute agreement, intraclass correlation coefficient (ICC) analysis was conducted. Results are presented in Table \ref{tab:agent_validation}.  All agents demonstrated acceptable mean scores and inter-rater reliability (ICC > 0.75 \cite{koo2016guideline}), supporting their integration into the multi-agent architecture.

\subsection{Interaction Interface}
We developed four web-based animated characters, as our simulated patients, to provide face-to-face user-simulated patient interactions (see \autoref{fig:animated_characters}). Users can access the website and interact with the simulated patient by text input. The simulated patient responds verbally through a text-to-speech synthesizer and visually with its response displayed at the bottom of the screen (see \autoref{fig:interaction}). This dual-modality output allows users to refer back to previous patient response without relying solely on their memory, addressing a challenge observed in our prior work, where participants struggled to recall the exact wording used by a virtual counselor (blinded for review). Upon accessing the interface, users must sign in and enter a unique session identifier to initiate the simulation. A brief tutorial screen outlining interaction guidelines is then presented.

The simulated patient is designed as a humanoid animated character, rendered in 3D on a computer screen, and situated within a simulated clinician's office (see \autoref{fig:interaction}). It communicates through spoken language and exhibits nonverbal behaviors such as gaze direction, posture, and hand movements, enhancing the realism of the interaction. These nonverbal behaviors are determined by two factors: the BEAT engine \cite{cassell2001beat} provides a baseline animation, and the Patient Response Agent within our Multi-Agent Architecture (see \autoref{fig:sim_patient}) uses GPT-4o to select specific nonverbal cues based on the current patient persona profile, session history, internal cognitive factors, and the user's most recent input. This process, similar to approaches in other systems \cite{lee2023developing, arjmand2024empathic}, seeks to align the nonverbal communication with the simulated patient's evolving internal state and the context of the conversation.  

User interactions are turn-based and text-driven: the agent speaks using a text-to-speech synthesizer, and the user types their response in free-text format. This approach circumvents potential inaccuracies introduced by automatic speech recognition. Users can terminate the conversation at any time by clicking an \textit{"End Conversation"} button, which triggers a transition to the evaluation dashboard.

\begin{figure*}[t]
\centering
\includegraphics[width=\textwidth]{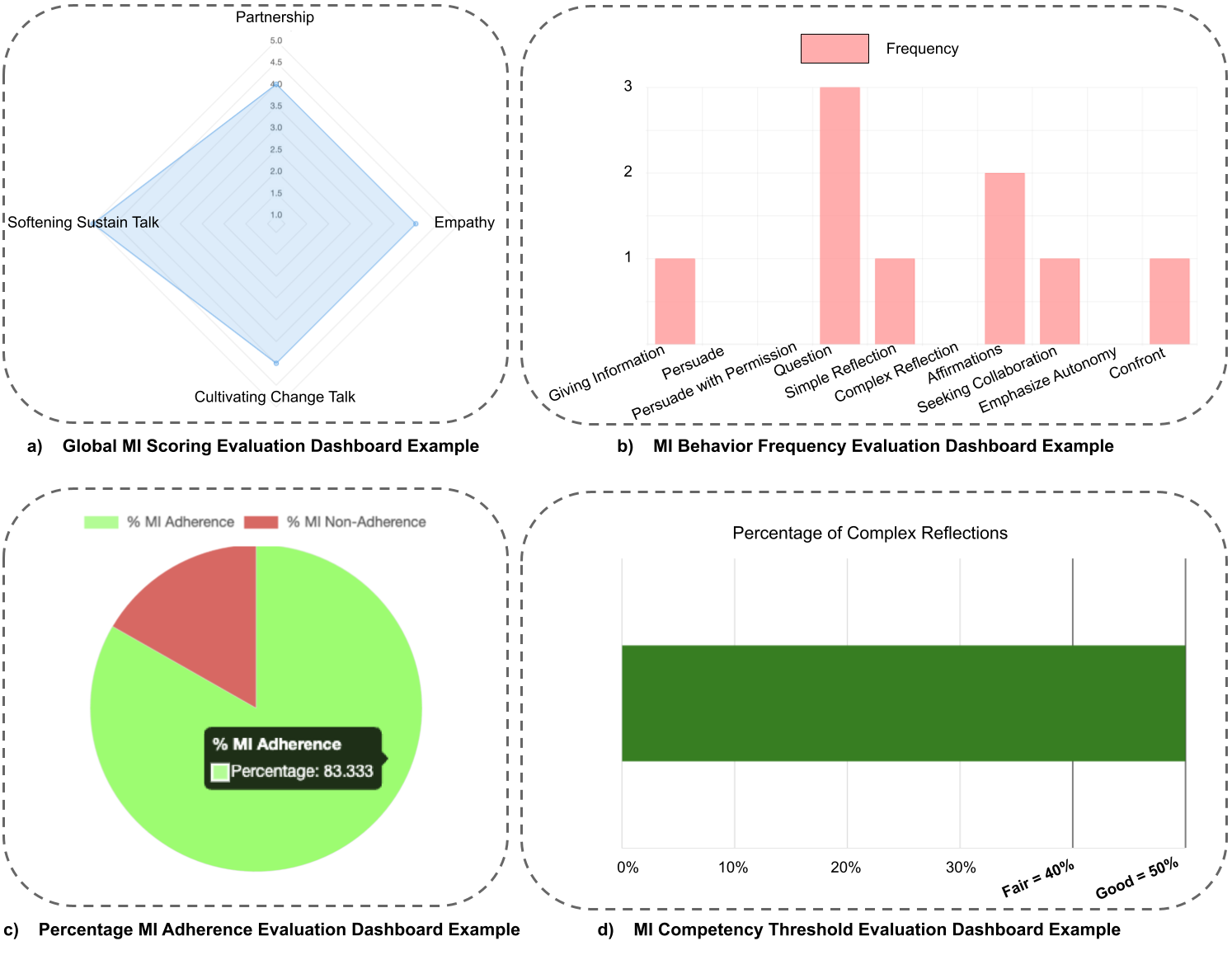}
\caption{Graphical Evaluation Dashboard Examples: This figure showcases example graphs from our MI skills evaluation dashboard. \textbf{a)} a radar chart visually represents scores on key Global MI measures ("Partnership", "Empathy", "Cultivating Change Talk", "Softening Sustain Talk"), rated on a scale of 1 to 5. \textbf{b)} a bar graph displaying the frequency of specific MI behavior codes used during a session. \textbf{c)} a pie chart that depicts the percentage of MI-adherent and non-adherent behaviors, highlighting adherence to MI principles. \textbf{d)} an example of one of four proficiency comparison bar graphs, such as Percentage of Complex Reflections, that depict "Fair" and "Good" proficiency thresholds. Additional details of the evaluation dashboard are available in Appendix \ref{apx:dashboard_example}.}
\label{fig:eval_dashboard_example}
\Description{This figure presents four examples of the graphical elements used in the evaluation dashboard: a radar chart summarizing global MI measures, a bar graph showing the frequency of different MI behavior codes used during a session, a pie chart showing the percentage of MI-adherent and non-adherent behaviors, and a bar graph comparing the user's score on a specific MI competency measure to established proficiency thresholds.}
\end{figure*}

\subsection{Evaluation Dashboard}
Upon concluding a session, users can access an evaluation dashboard to review their MI performance. This web-based dashboard, accessible after signing in and selecting the desired session, provides a multifaceted assessment that includes evaluation summaries, numerical and graphical measures, and a detailed transcript. The performance metrics presented on the dashboard were all taken from the MITI coding manual \cite{moyers2016motivational}. These metrics are calculated using several LLM agents within the SimPatient system. Critically, the global MI scores, MI behaviors, cognitive factor changes presented on the dashboard are accompanied by justifications generated by their corresponding LLM agents within the SimPatient system. The following sections detail the dashboard modules available to SimPatient users.

\subsubsection{Session Summary}
The first module presents a concise, paragraph-long summary of user session performance. This summary highlights strengths, areas that need improvement, and recommendations for future practice.

\subsubsection{MI Description}
The second module provides a description of what MI is, offering a concise refresher on its core principles and the importance of \textit{"change-talk"} and \textit{"sustain-talk"}. This description is based on the latest edition of MI by its authors \citeauthor{miller2023motivational}: \textit{"Motivational Interviewing
Fourth Edition Helping People Change and Grow"} \cite{miller2023motivational}.

\subsubsection{Global MI Scoring}
The third module showcases the user's global MI scores on four key dimensions: "Partnership", "Empathy", "Cultivating Change Talk", and "Softening Sustain Talk". Each dimension is rated on a scale from 1 (low) to 5 (high), visually represented on a radar chart to facilitate rapid assessment of strengths and areas for improvement (see a) in \autoref{fig:eval_dashboard_example}). Short descriptions of each measure are provided along with individualized justifications for the assigned scores. For example, a user receiving a score of 3 out of 5 on "Softening Sustain Talk" may see the following rationale: \textit{"The counselor occasionally addresses sustain talk but at times lingers in it, especially when the patient expresses doubts (e.g., `It sounds good in theory, but honestly I just cannot see anything that would give me that same relief.'). The counselor does attempt to shift the focus towards positivity, but could enhance the practice by further minimizing engagement with sustain talk"}. Users can hover over each data point on the radar chart to reveal the corresponding global score value. This module, with its descriptive explanations and score justifications, directly addresses stakeholder feedback from our formative study, addressing the need for transparency and personalized feedback.

\subsubsection{MI Behavior Frequency}
The fourth module visualizes the frequency of each MI behavior code used by the user during the session. A bar graph (see b) in \autoref{fig:eval_dashboard_example}) shows the behavior counts for giving information, persuading, persuading with permission, question, simple reflection, complex reflection, affirmation, seeking collaboration, emphasizing autonomy, and confront. Concise descriptions accompanying each behavior code provide convenient reminders for users. Hovering over each bar reveals the corresponding frequency count and label.

\subsubsection{Percentage of MI Adherence}
The fifth module utilizes a pie chart to present the percentage of MI adherence and MI non-adherence (see c) in \autoref{fig:eval_dashboard_example}). These percentages are calculated on the basis of the proportion of MI-adherent and MI-non-adherent behavior codes relative to the total code count. The MI-adherent count includes the MI behavior codes of affirmations, seeking collaboration, and emphasizing autonomy, while the MI-non-adherent count includes persuade and confront. Users can hover over each segment of the pie chart to view the corresponding percentage and label.

\subsubsection{MI Competency Thresholds}
The sixth module utilizes four bar graphs (see d) in \autoref{fig:eval_dashboard_example}) to represent a user's MI proficiency and competency based on thresholds derived from the MITI coding manual \cite{moyers2016motivational}. The MI Relational Global Score, calculated by averaging the "Empathy" and "Partnership" global ratings, provides an indication of the extent to which the counselor created a collaborative alliance and demonstrated understanding of the client's perspective. The MI Technical Global Score, calculated by averaging the "Softening Sustain Talk" and "Cultivating Change Talk" global ratings, reflects the degree to which the counselor guided the conversation toward change while strategically responding to client language. The Percentage of Complex Reflections, is derived by dividing the number of complex reflections by the total number of reflections. It indicates the proportion of reflections that went beyond the client's initial statements to add meaning or emphasis. Finally, the Reflection-to-Question Ratio is calculated by dividing the total number of reflections by the total number of questions, which provides insight into the balance between the counselor's use of active listening and directing the session. Each bar graph allows the user to compare their score against the "Fair" and "Good" proficiency thresholds. Lastly, a color-coding feature, implemented based on user suggestions during formative study, aims to enhance interpretability. Scores meeting or exceeding the "Good" threshold are colored dark green, those within the "Fair" range are colored light green, and those below the "Fair" threshold are colored yellow.

\begin{figure}[htb]
\centering
\includegraphics[width=\columnwidth]{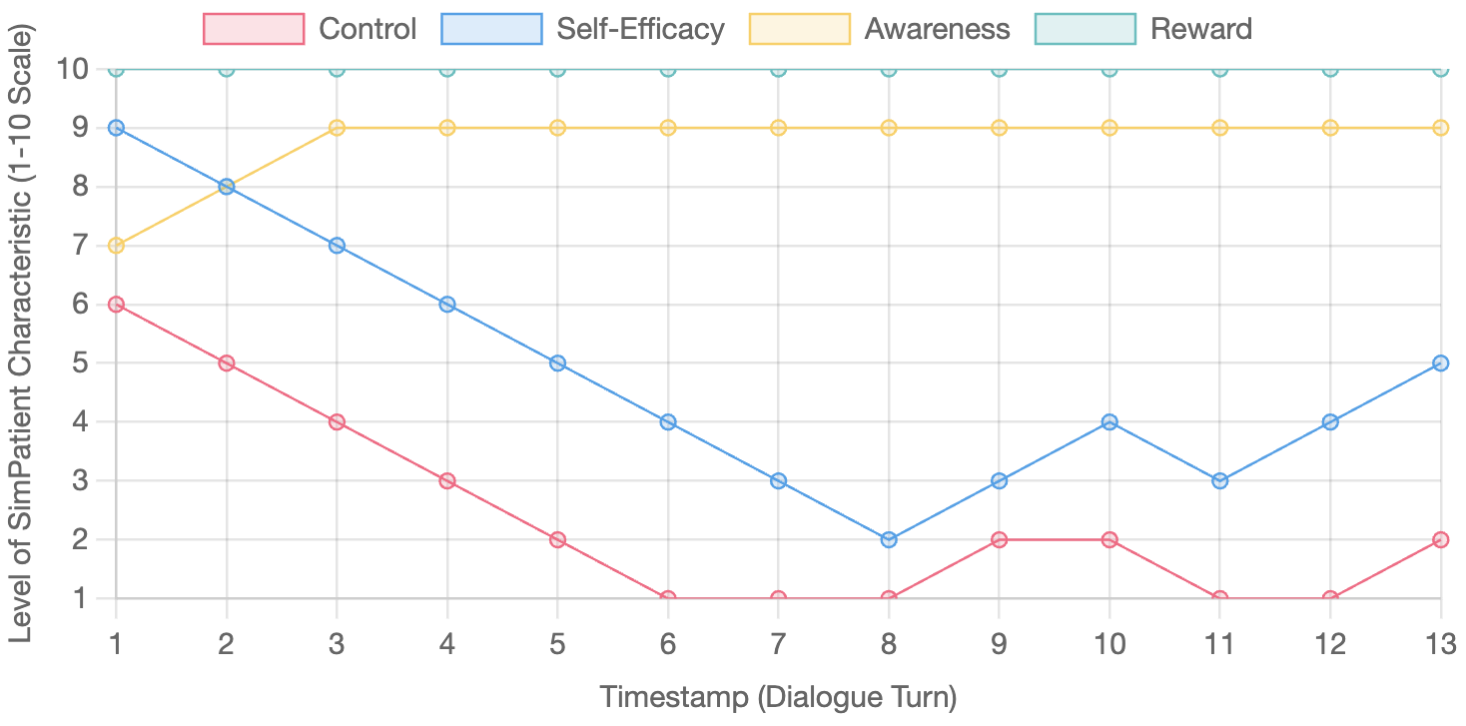}
\caption{Dynamic Cognitive Factors Graph}
\label{fig:cognitive_factors_graph}
\Description{This figure exemplifies how SimPatient visualizes the changes in the simulated patient's internal states during the conversation. It shows a line graph where the x-axis represents the progression of the conversation (either by timestamps or dialogue turns), and the y-axis represents the score of the four dynamic cognitive factors: Control, Self-Efficacy, Awareness, and Reward.}
\end{figure}

\subsubsection{Dynamic Cognitive Factors}
The seventh module visualizes the fluctuations of our four dynamic cognitive factors throughout the session - Control, Self-Efficacy, Awareness, and Reward - providing insight into the internal state of the simulated patient. Each cognitive factor is represented on a 1 to 10 scale and updated at every patient response to a user's most recent utterance. The module provides a brief explanation of each cognitive factor and displays their values as they evolve over the course of the session, plotted on a line graph (see \autoref{fig:cognitive_factors_graph}). The x-axis of the line graph corresponds to timestamps or dialogue turn numbers from the session transcript, allowing users to correlate specific user utterances and patient responses with changes in the simulated patient's internal state. For example, consistent use of an MI-adherent skill such as affirmations may lead to an increase in the patient's self-efficacy, while confrontation attempts might trigger a decrease. This visual representation of the dynamic cognitive factors, driven by stakeholder feedback during our formative study, aims to enhance the counselor's understanding of the simulated patient's behavior and responsiveness to different communication styles.

\subsubsection{Session Transcript}
The eighth and final module presents the complete session transcript, beginning with the user's initial utterance and concluding with the patient's last response before the conversation's termination. The transcript is organized chronologically by user utterance and patient response.

Each user's typed utterance is displayed, followed by a bulleted list of assigned MI behavior codes and justifications for those codes. Similarly, each patient response section displays the generated response along with a bulleted list of current patient cognitive factors and justifications for their values or changes. The inclusion of these justifications comes directly from findings in formative study, where both professional and student counselors expressed a desire to understand the rationale behind the coding of user utterances and the fluctuations in cognitive factors, particularly self-efficacy.

\begin{figure*}[htb]
\centering
\includegraphics[width=\textwidth]{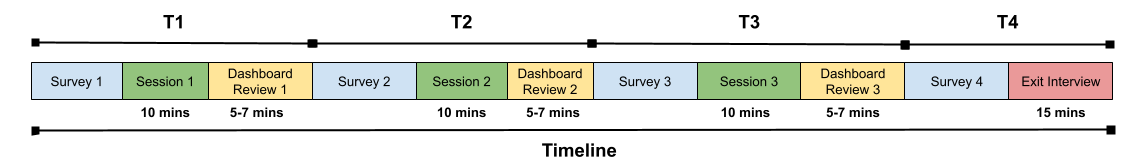}
\caption{A flowchart of the Motivational Interviewing Training Study}
\label{fig:flowchart}
\Description{This flowchart illustrates the study design and procedure. It details the different stages of the study, including the pre-session and post-session surveys (T1-T4) and the exit interview, as well as the points at which participants interacted with the SimPatient training system.}
\end{figure*}

\section{Motivational Interviewing Training Study}
We conducted a preliminary evaluation of SimPatient using a  within-subjects repeated measures design \cite{sullivan2008repeated} to evaluate pre-post changes in MI confidence across each of the multiple sessions, and assess its usability and user satisfaction among both professional and student counselors. We felt that experts could contribute professional insights into the validity and relevance of the simulated patient’s behavior and feedback, while students could offer a beginner's perspective on usability and learning potential. 

Participants each evaluated one of eleven possible patient personas selected at random, representing a range of ages, genders, ethnicities, occupations, and Myers-Briggs personality types \cite{briggs1976myers},  embodied by four animated character models (see \autoref{fig:animated_characters}. Each participant was randomly assigned a persona using block randomization, where they interacted with the same persona across all three sessions (\autoref{fig:flowchart}).  
In addition, the four cognitive factors were randomly assigned initial values at the start of the first session, reflecting different mindsets of patients about the reduction of alcohol intake. 

Participants engaged in three 10-minute sessions.
Each session ended with a 5 to 7-minute review of the evaluation dashboard. Pre- and post-session surveys were used after every session to assess participants' self-efficacy with MI, allowing us to track their potential confidence shifts over time. After completing all sessions and surveys, participants engaged in a 15-minute semi-structured exit interview. An overall flowchart of the study is shown in \autoref{fig:flowchart}. This study adopted the same participant inclusion criteria employed in the formative study. The 90-minute study was approved by our institution's IRB and the participants were compensated for their time.

\subsection{Measures}
The following is a detailed breakdown of all the measures that we collected from the participants for our MI training study. 

\textbf{MI Self-Efficacy}
The Motivational Interviewing Confidence Survey \cite{larson2021measuring} was used to assess participants' confidence levels in using MI techniques. The survey comprises 24 items, each rated on a scale from 0 (Cannot do at all) to 10 (Highly certain can do).

\textbf{System Usability, Utility, Impact, and Patient Realism}
Participants provided feedback during a post-training survey (T4, see \autoref{fig:flowchart}) in four key areas: 
\begin{itemize}
    \item \textbf{Training Impact} was measured to evaluate how well the SimPatient system was performing in its training functionality, unlike the MI Self-Efficacy measure, which gauges participants' general self-perceived confidence in MI. Training impact was assessed using a single item: "How much did the training help you improve your motivational interviewing skills?", with a 10-point scale ranging from 1 ("not at all") to 10 ("completely").
    \item \textbf{Evaluation Dashboard \& Metrics Utility} was assessed using another 10-point single item "How useful was the evaluation dashboard and the metrics provided after each session?"
    \item \textbf{Patient Change Realism} was assessed using "How realistic were the changes in the patients’ cognitive factors observed during the training, such as self-efficacy?"
\end{itemize}
The overall usability of the simulated patient system was evaluated using the System Usability Scale (SUS) \cite{brooke1996sus}.

\textbf{Patient Persona Fidelity}
Future work will incorporate user selection of patient personas to accommodate the needs of health professionals who specialize in working with specific demographics, wish to broaden their skills across diverse patient populations or reduce their implicit biases \cite{chapman2013physicians}. To lay the groundwork for this functionality and assess the fidelity of the simulated patient's portrayal of assigned personas, a post-training survey (T4, see \autoref{fig:flowchart}) asked participants to identify characteristics of the simulated patient with whom they interacted. These characteristics included gender (male, female, or non-binary/third-gender), age (18, 28, 38, 48, 58, or 68), ethnicity (White, Asian, Black or African American, Hispanic, Latinx or Spanish Origin, or Middle Eastern or North African), occupation (student, cashier, nurse, cook/chef, retail salesperson), and personality type (Myers–Briggs personality types \cite{briggs1976myers}). The responses of the participants were then compared with the actual characteristics assigned. 
A binomial test for significance was conducted for each characteristic, comparing the observed proportion to the proportion expected by chance (e.g., 33.33\% if participants were guessing randomly on gender: male, female, or non-binary/third gender), to determine if the multi-agent LLM architecture successfully conveyed the intended persona.

\begin{figure}[htb]
\centering
\includegraphics[width=\columnwidth]{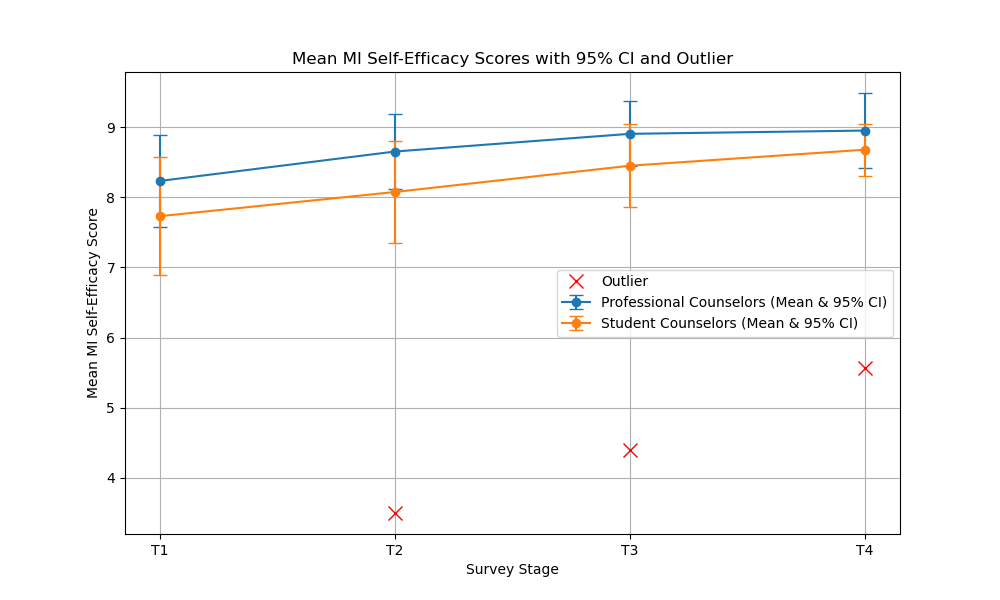}
\caption{MI Self-Efficacy Scores by Participant Class \& Survey Stage}
\label{fig:mi_self_efficacy}
\Description{MI self-efficacy scores of the participants over time, showing change from  initial assessment (T1) to final assessment (T4), for professional  and student counselors.}
\end{figure} 

\subsection{Results}
\subsubsection{Participants}
Our study initially included 19 participants: 10 professional counselors and 9 student counselors. The professional counselors (7 female, 2 male, and 1 demi-girl) ranged in age from 27 to 46 years (M = 33.6, SD = 6.31). The most common occupations within this group were clinical psychologist (n=2), registered dietitian (n=3), and school social worker (n=2). The student counselors (6 female, 1 male, 1 non-binary, and 1 undisclosed) ranged in age from 23 to 32 years (M = 26.22, SD = 3.37). The majority of student counselors (n=5) were pursuing master's or doctorate degrees in counseling psychology. Due to a networking issue, data from one student counselor was removed as they were unable to complete the entire 90-minute study. The final sample for analysis consisted of 18 participants.

\subsubsection{MI Self-Efficacy}
Our primary outcome was the change in participants' MI self-efficacy across the three sessions. One outlier, identified due to Z-scores between -5.5 and -6.4, was removed from the analysis. Overall, MI self-efficacy scores increased from a mean of 8.08 (SD = 0.95) at T1 to 8.82 (SD = 0.59) at T4. A repeated measures ANOVA showed a significant effect of time for all participants, F(3, 48) = 15.56, p < 0.001. Post-hoc pairwise t-tests with a Bonferroni correction revealed significant increases between T1 and T2, T1 and T3, T1 and T4, T2 and T3, and T2 and T4.

\begin{table*}[h!]
\centering
\begin{tabular}{|l|c|c|c|c|}
\hline
\textbf{} & \textbf{Train. Impact} & \textbf{Evaluation Utility} & \textbf{Patient Change Realism} & \textbf{System Usability} \\ \hline
\textbf{Overall}       & 7.97 (1.33)  & 8.83 (1.34)  & 7.94 (1.74) & 88.1 (8.60) \\ \hline
\textbf{Professional Counselors}       & 7.9 (1.66)   & 8.4 (1.58)   & 7.5 (2.18)  & 89.0 (7.19) \\ \hline
\textbf{Student Counselors}      & 8.06 (1.00)   & 9.34 (0.97)   & 8.44 (0.98)  & 86.88 (10.50)\\ \hline
\end{tabular}
\caption{Mean (SD) for Training Impact, Evaluation Utility, Patient Change Realism, and System Usability. Measures were captured during a post-training survey at T4.}
\label{tab:misc_scores}
\Description{Participant ratings of training impact, evaluation utility, patient change realism, and system usability, collected during the post-training survey at T4, for professional and counselors separately and combined.}
\end{table*}

\subsubsection{System Usability, Utility, Impact, and Patient Realism.}
The overall mean ratings for Training Impact were 7.97 (SD = 1.33), 8.83 (SD = 1.34) for Evaluation Dashboard \& Metrics Utility, and 7.94 (SD = 1.74) for Patient Change Realism. 
The SUS score was 88.1 (SD = 8.60), reflecting a well-performing system with an "Excellent" rating \cite{bangor2008empirical}. \autoref{tab:misc_scores} provides a detailed breakdown of these scores between professional and student counselors.

\subsubsection{Patient Persona Fidelity}
There was significant agreement between the attributes of the assigned persona and the perceived characteristics of the persona on gender (94\% accuracy, p < 0.01) and occupation (72\% accuracy, p < 0.01), suggesting that the system effectively conveyed these specific persona characteristics. However, no significant agreement was found for age, ethnicity, or personality type.

\subsubsection{Qualitative Results}
Our thematic analysis of the semi-structured interview transcripts revealed several findings about our system design and interaction.

\textbf{Conversational Fidelity and Realism}. Participants generally found the patient responses to be sensible, contextually appropriate, and similar to those they encounter in real-world interactions. One student counselor remarked that the simulated patient provided responses "\textit{in a very sensible manner...it was like I was talking to a person.} [P4, student counselor]" This sense of lifelikeness stemmed from the simulated patient's ability to generate human-like responses. However, some participants, such as one student counselor, found certain responses to be somewhat "\textit{machine-like} [P5,  student counselor]" particularly in situations where the simulated patient exhibited limited assertiveness or failed to convincingly "lie" about its actions between sessions. The use of colloquial terms and informal language contributed to the perceived realism, as one student counselor noted that the simulated patient "\textit{using colloquial terms, using the word "like" in sentences definitely helped it make it seem a bit more realistic.} [P10,  student counselor" However, the robotic nature of the text-to-speech synthesizer was viewed as a drawback by some, including P10, who suggested that "\textit{having a bit more intonations would be good for helping it come across as more lifelike} [P10,  student counselor]".

\textbf{Simulating Resistance and Ambivalence}.
While many participants praised the conversational abilities of the simulated patient, several expressed a desire for greater resistance and ambivalence towards change. One student counselor reflecting on the overall eagerness of the simulated patient, suggested "\textit{if you guys can adjust...like client ambivalence levels...I think depending on the reasons for a client coming in...they might be a lot more resistant.}" [P8,  student counselor]. This desire for increased resistance reflects the reality that many individuals struggling with alcohol misuse exhibit significant ambivalence towards change \cite{room1976ambivalence}. Another student counselor echoed this sentiment, noting, "\textit{The number of relapses...it’s much higher than what happened with this patient} [P9,  student counselor]". 

Several participants, including P7 and P11, explicitly requested the ability to interact with patients exhibiting varying degrees of resistance. P7 suggested, "\textit{maybe the perfection of the client and needing a little bit more challenge to it since a lot of the MI skills come in around resistance and ambivalence} [P7,  professional counselor]" while P11 advocated for "\textit{being able to dig into some really hardcore resistance} [P11,  professional counselor]".

\textbf{Perceived Realism of Dynamic Cognitive Factors}.
The inclusion of dynamic cognitive factors, designed to fluctuate based on user interactions, was widely viewed as a valuable training feature. Participants appreciated the ability to witness the impact of their MI skills on the simulated patient's internal state, a dimension often obscured in real-world interactions. One student counselor, reflecting on the efficacy of these visualizations, stated, "\textit{so those changes in the responses, they were actually reflecting on the conversation} [P12,  student counselor]" while one student counselor acknowledged, "\textit{It was neat to actually track the patient's...experience too} [P16,  student counselor]".

However, several participants noted that the specific fluctuations in the "reward" were questionable. One professional counselor observed that \textit{"the rigidity of the reward...I think was interesting} [P7,  professional counselor]" while another professional counselor acknowledged, "\textit{I don’t know if reward really could be affected in that person} [P14,  professional counselor]".

\textbf{Utility of Evaluation Dashboard \& Metrics}.
The dashboard's  evaluation metrics were generally well-received, with participants finding them valuable for understanding their performance and identifying areas for improvement. One student counselor appreciated the visual clarity, stating, "\textit{I liked the graphs...they were really self-explanatory}" [P4,  student counselor]. Another participant, referencing the bar graph of MI behavior code frequency, noted that seeing their low score on affirmations prompted them to "\textit{change my language a little bit.}" [P2,  professional counselor].

The inclusion of competency thresholds, derived from the MITI coding manual, was particularly valued for benchmarking performance against established MI proficiency standards. A professional counselor, reflecting on their progress in the reflection-to-question ratio, stated: "\textit{I think I probably was a 0.3 on the first two, and I got it up to a 0.5...in terms of training tool, how could I get that up as the user?}" [P2,  professional counselor]. Participants found that comparative feedback motivated improvement by providing a clear target to aim for. However, while the dashboard effectively revealed areas for improvement, some expressed a desire for more concrete recommendations. One participant, referring to their feedback on missed opportunities for deeper exploration, suggested including "\textit{an example of that, oh, ask more open-ended questions, or when this person said this, here's how you could have dove deeper, a bit more specificity with how somebody can improve}" [P6, female professional counselor]. This desire for concrete examples was echoed by another participant, who noted, "\textit{I'm one of those people that needs an example, so I couldn't figure it out.}" [P14, professional counselor].

A key strength highlighted by many participants was the inclusion of individualized justifications for the assigned scores. One student counselor  remarked, "\textit{It [the detailed feedback] was really helpful. It made a lot of sense}" [P5,  student counselor]. However, there were calls for enhanced clarity and accessibility in presenting the scoring rubric. One participant suggested that "\textit{it would've helped to have that the scale follow me so that I would be able to reference...what does the Q mean... that way I didn't have to keep going back and forth.}" [P3,  professional counselor].

The provision of a complete annotated transcript was also highly praised by participants, who found it to be a valuable tool for reflection and learning. One student counselor, initially skeptical about the transcript's utility, stated, "\textit{I actually really liked the transcript…it was more helpful than I thought it would be originally}" [P19,  student counselor]" emphasizing that they "\textit{spent the most time in that five-minute stretch looking at my actual responses and how it was coding things}" [P19,  student counselor]". This sentiment was echoed by a professional counselor, who found it "\textit{very helpful...the examples of those characteristics in the transcript...how you responded or how you elicited that}" [P11,  professional counselor].

Furthermore, participants perceived that the organization of the transcript into user utterances and patient responses, with accompanying justifications for codes and cognitive factor values, promoted a deeper understanding of the interplay between communication styles and patient reactions. For instance, reflecting on the changes to the simulated patients' cognitive factors within the transcript, one professional counselor noted: "\textit{When something would decrease or something would increase, I think that kind of reaffirmed like, oh, whatever I sent there, that was good. Do that again}" [P7,  professional counselor]".

\textbf{Accuracy of Metrics}.
Participants generally expressed confidence in the accuracy of the evaluation metrics, finding them to be consistent with their perceptions of the interactions. One professional counselor, reflecting on the overall alignment between their performance and the dashboard feedback, remarked, "\textit{it was accurate because I don't think it was too much statistically different}" [P2,  professional counselor]. Similarly, another professional counselor, when asked about the accuracy, simply stated, "\textit{Yeah, I feel like it did make sense...I don't feel like there was anything that was inaccurate.}" [P3,  professional counselor].
This participant did acknowledge the potential for subtle nuances in communication to go unnoticed, suggesting that "\textit{There might be some things here and there that I didn't notice that might've been a bit more nuanced.}" [P11,  professional counselor]. Another professional counselor, echoing the sentiment of the perceived MI assessment accuracy, stated, "\textit{Yeah. I feel like it was reflected pretty well on the quantitative dashboard.}" [P6,  professional counselor]. Furthermore, participants found that the system accurately captured and reflected deliberate shifts in communication style. One professional counselor, who intentionally adopted a less MI-consistent approach in one session, observed, "\textit{I did do more non motivational interviewing in a second one to see how direct it would pick up on directly focused, more rational emotive thoughts. And it did, it totally picked up on it. My whole green graph turned to partially red}" [P14,  professional counselor]. This ability to detect and display variations in communication style reinforced participants' belief in the system's accuracy.

\textbf{SimPatient in Curricula}.
Participants overwhelmingly envisioned broad applications for SimPatient across diverse educational and training contexts. One professional counselor emphasized its potential as a scalable training tool, stating, "\textit{I think it has a chance to be a home run because it's a great training tool to use in mass for a provider to do motivational interviewing.}" [P10,  professional counselor]. This sentiment was echoed by others who saw its relevance for a wide range of professional counselors, including mental health practitioners, coaches, and even those working with individuals struggling with gambling addiction [P2, P3]. Furthermore, participants highlighted its potential to improve upon existing training methods. One professional counselor, reflecting on their own training experiences using role-playing with colleagues, shared, "\textit{To actually have [SimPatient] in those sessions...rather than practice on each other...I think this would be invaluable.}" [P3,  professional counselor]. The ability to practice skills in a safe and controlled environment, without the pressure of real-world consequences, was viewed as a significant advantage.

Many participants drew direct comparisons between SimPatient and traditional role-playing exercises, noting its superiority in providing structured feedback and objective evaluation. One student counselor, contrasting SimPatient with the standard practice of recording and reviewing sessions with peers, remarked, "\textit{being able to see it written out and highlighted...And then also on the client side, being able to identify those pieces with them as well was really cool. So it just really gave me a very different perspective.}" [P11,  professional counselor]. This sentiment underscores the value of the detailed, multi-faceted feedback provided by SimPatient, which surpasses the often-limited insights gained from traditional peer evaluations.

\textbf{Typing vs. Speaking Modalities}.
The typing interface, while not mirroring real-world consultations, provided notable advantages for skill development. Participants found that typing encouraged more deliberate and thoughtful responses. One professional counselor acknowledged, "\textit{If I was in person, I would be like, oh, I'm just going to blurt out this question. But with the typing, I was like, oh, hold on. Let me slow down and really think about what I want to say.}" [P6, female professional counselor]. This capacity for reflection aligns with the core principles of motivational interviewing, promoting careful consideration of both content and delivery. Another participant, highlighting the benefits of editing, stated, "\textit{I noticed myself going back and changing things a couple times...It's nice. It gives you a different perspective versus if you're just talking and you don't have the opportunity to go back and delete something.}" [P11,  professional counselor].

However, participants also recognized the limitations of a solely text-based system, particularly for more experienced practitioners. One professional counselor suggested that "\textit{somebody more experienced would benefit from [speaking]...you have to stay really, really calm and collected... It's so easy to do that if you're texting...versus if you're speaking and it needs to be back and forth.}" [P9,  professional counselor]. This feedback suggests that the addition of a speech-based interface could provide valuable training in managing the real-time pressures of spoken interaction, such as maintaining composure, thinking on one's feet, and responding to nuanced verbal and nonverbal cues. Many suggested that offering both modalities, potentially controlled by an administrator, would be ideal, allowing for a tailored and comprehensive training experience [P9, P10, P11, P19].

\textbf{Dashboard Preferences}
Feedback revealed diverse preferences across the dashboard components, including the session summary, visualizations, and session transcript. The pie chart depicting MI adherence consistently received positive feedback, with participants appreciating its clear, concise representation of MI-consistent and non-MI-consistent responses as helpful for training.

Other modules elicited more varied reactions, though none were widely disliked. The line graph illustrating changes in patient characteristics (self-efficacy, control, awareness, reward) sparked both appreciation for its visualization of interaction impact and concerns regarding interpretability and overall usefulness compared to other graphs. The "reward" characteristic, in particular, drew questions regarding its representation and realism.

A recurring recommendation was to divide the dashboard into selectable sections to improve navigation and avoid information overload. 

\section{Discussion}
\subsection{RQ1: Design Features for MI Training Systems}
Our formative study captured a range of preferences regarding the design of an MI training and skill assessment system for both novice trainees and experienced professionals. Participants, particularly professionals, emphasized the importance of evaluation measures aligned with established MI principles \cite{miller2023motivational, moyers2016motivational}. Their desire for diverse visualizations—graphical, numerical, and transcript-based—reflects the importance of accommodating different learning styles in user-centered design, catering to both visual-nonverbal (e.g., graphs and numerical) and visual-verbal (e.g., textual feedback and transcripts) learners  \cite{jensen2024automated}. 

Participants expressed a desire for more granular, actionable feedback alongside the scores provided by the agents. While the chain-of-thought prompting used in our multi-agent architecture generated well-received feedback, participants wanted more specific guidance on implementing the recommendations. For example, a suggestion to "delve deeper into the patient's unexpressed thoughts" could be augmented with a concrete example like, "Try asking the patient, 'You mentioned feeling anxious. What thoughts are going through your mind when you feel that way?.'" Given the success of LLMs and chain-of-thought prompting in providing personalized feedback in our system and as demonstrated in other related work \cite{yang2024chain, cohn2024chain}, this approach, coupled with further exploration of prompt engineering for more granular feedback, holds promise for other social skills training systems as well. 

\subsection{RQ2: Perception \& Response to SimPatient}
Participants rated the SimPatient system highly in terms of usability and effectiveness, where they suggested its use in real-world application in various training contexts, such as continuing education for practicing counselors, initial skills training for students in counseling programs, and potentially even in broader healthcare education settings where communication skills are emphasized. Additionally, we found that MI self-efficacy among both student and professional counselors increased as they used the system across multiple sessions. Thus highlighting the system's potential for training across experience levels.

The perceived realism of patient responses was a recurring strength of SimPatient, particularly among students, underscoring the value of realistic training environments \cite{khan2011simulation, good2003patient}. However, participants also highlighted a critical area for improvement in LLM-driven simulated patients: the need for greater resistance and ambivalence. As noted in related work, LLMs, particularly those trained with reinforcement learning from human feedback, often exhibit a positivity bias, tending towards agreeableness and compliance \cite{perez2022discovering}. This inherent tendency poses a challenge for simulating the full spectrum of patient behaviors, especially resistance and deception, which are crucial for robust counselor training \cite{lee2024llms, petrov2024limited, mieleszczenko2024dark}. Our participants' feedback echoes this limitation, noting that the simulated patients primarily reflected motivated individuals ready for change. 

\subsection{RQ3: Patient Persona Fidelity}
Our exploration of LLM-driven patient persona simulation yielded mixed results. While participants accurately identified gender and occupation, suggesting the LLMs captured these simpler demographic attributes, the inconsistent recognition of more complex traits like ethnicity and personality type highlights ongoing challenges in persona representation using LLMs \cite{lee2024llms, petrov2024limited, mieleszczenko2024dark}. Improving the ability of LLMs to embody nuanced and multifaceted personas is an ongoing research problem and is essential for creating more effective and inclusive training simulations.

\subsection{RQ4: Dynamic Cognitive Factors Fidelity}
Feedback on the dynamic visualization of cognitive factors, especially self-efficacy, was overwhelmingly positive, particularly among student counselors who found the line graph illustrating these changes to be highly beneficial for learning. They perceived the fluctuations as realistic given the conversation flow and valued the insights it provided into interaction dynamics and the impact of their MI techniques. This provides preliminary evidence for the value of incorporating cognitive models alongside LLMs in various training environments where understanding the impact of one's actions on others' cognitive and emotional states is crucial, such as those focusing on interpersonal skills \cite{ladd1983cognitive}, healthcare communication \cite{radwin1995knowing, adams2012should}, and conflict resolution \cite{smith2024impact}.

\subsection{Limitations \& Future Work} 
Our research provides  insights into the design and evaluation of LLM-powered MI training systems; however, certain limitations warrant acknowledgment. First, while our within-subjects repeated measures design offered valuable insights into the SimPatient system's effects, the inclusion of a control group is required for a rigorous assessment. Future work using a randomized controlled trial will compare the SimPatient system to an active control group receiving either traditional didactic MI training or assigned reading materials on MI. Second, the study's concentration on alcohol misuse may restrict the generalizability of findings to other counseling contexts. 
In addition, larger differences in patient resistance, as well as possible forms of "deception", may help broaden the learning capabilities of training systems. Future research should investigate the system's adaptability and efficacy in addressing diverse client populations and various levels of patient resistance. Third, while self-efficacy serves as a valuable surrogate measure for MI skill acquisition, future studies should incorporate more objective assessments of MI proficiency, such as pre- and post-training evaluations with standardized patients or analyses of real-world counseling sessions. Fourth, it is important to acknowledge that this study did not formally validate the SimPatient system's accuracy in reflecting intended persona characteristics and cognitive state fluctuations. Our evaluation focused primarily on counselor perceptions of fidelity, providing valuable insights into user experience and perceived realism, but not guaranteeing objective accuracy. Furthermore, potential biases within the language model, including gender and racial stereotypes, were not addressed in this study's design or analysis. This is a crucial area for future research, particularly given increasing evidence of bias in agent personas \cite{gupta2023bias, liu2024evaluating}. Future work should incorporate more rigorous validation methods, such as structured expert evaluations of simulated patient behavior, comparisons with real patient data, and evaluations by a socio-demographically diverse group of human raters to assess persona fidelity and identify potential biases. Fifth, as quantitative data on preferences for individual dashboard modules were not collected, our results reflect the holistic feedback on the dashboard. Future work will investigate individual module preferences quantitatively to further refine dashboard design and optimize user experience. Lastly, reliance on a text-based interface, though advantageous for fostering deliberate responses, may not fully capture the real-time pressures and nonverbal communication nuances inherent in face-to-face counseling. Integrating speech-based interaction and potentially incorporating virtual reality environments could enhance the ecological validity of the training experience.

\section{Conclusion}
This work explores the design and evaluation of SimPatient, an LLM-powered training system for motivational interviewing skills, such as empathy, focusing on alcohol misuse. Our formative study revealed stakeholder preferences for detailed, multimodal feedback, and realistic patient interactions. The subsequent MI Training Study demonstrated high usability and satisfaction with the system across expertise level, particularly its comprehensive evaluation dashboard and the inclusion of dynamic cognitive factors that realistically reflect changes in patient states during counseling sessions.

SimPatient successfully addresses the limitations of traditional counselor training methods by providing frequent, utterance-level feedback and a safe, accessible practice environment, leading to increased MI self-efficacy in student and professional counselors. Our findings suggest promising applications for this LLM-powered approach across various social skills training contexts and experience levels, paving the way for more effective and efficient skill development in a variety of domains.

\bibliographystyle{ACM-Reference-Format}
\bibliography{bibliography}

\appendix
\section{\\Appendix A – Evaluation Dashboard Example}
\label{apx:dashboard_example}

Below is an example screen-capture of the entire evaluation dashboard that participants in the MI training study saw during their first session with SimPatient.

\begin{figure*}[htb]
\centering
\includegraphics[width=\textwidth]{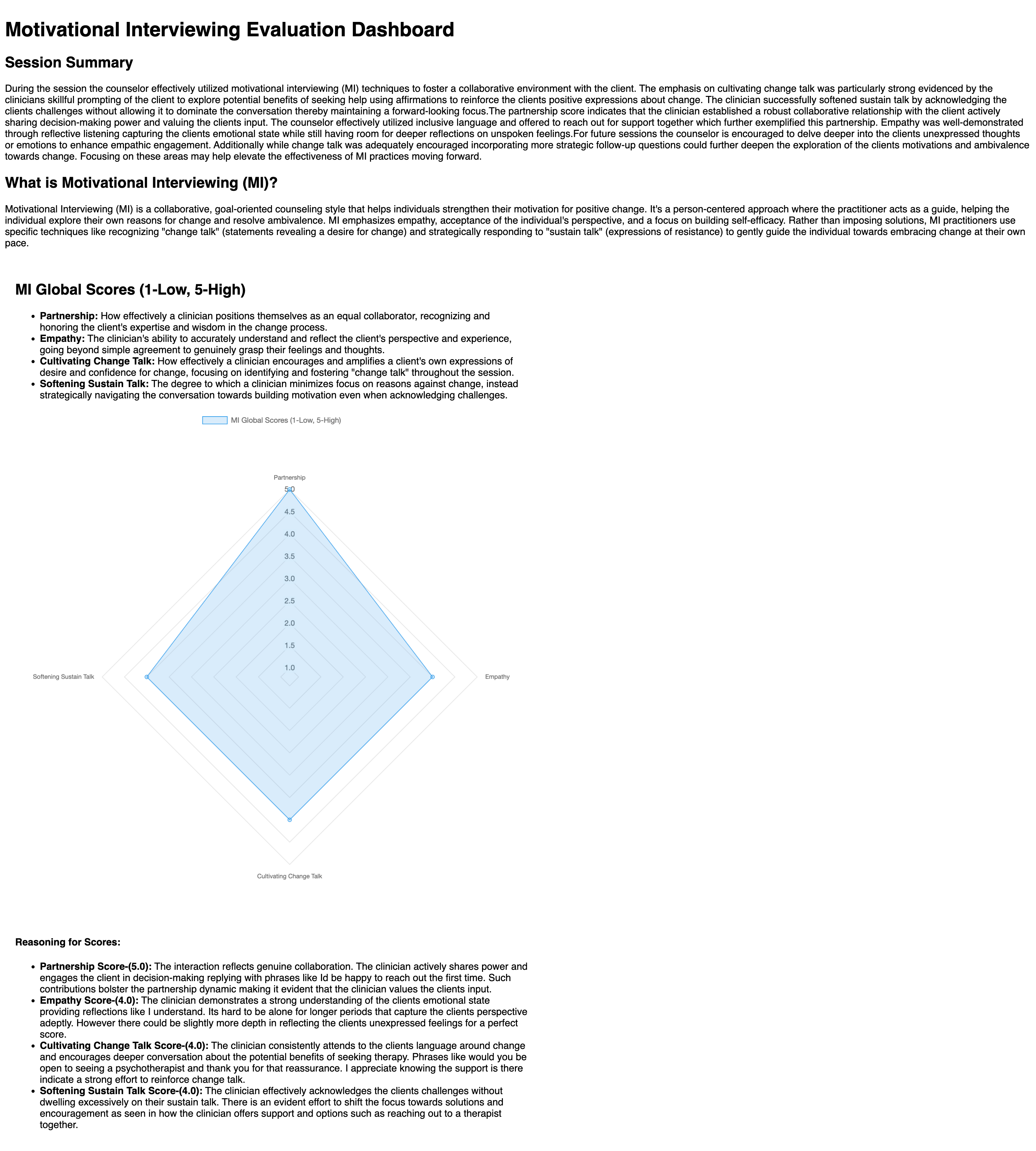}
\end{figure*} 
\begin{figure*}[htb]
\centering
\includegraphics[width=\textwidth]{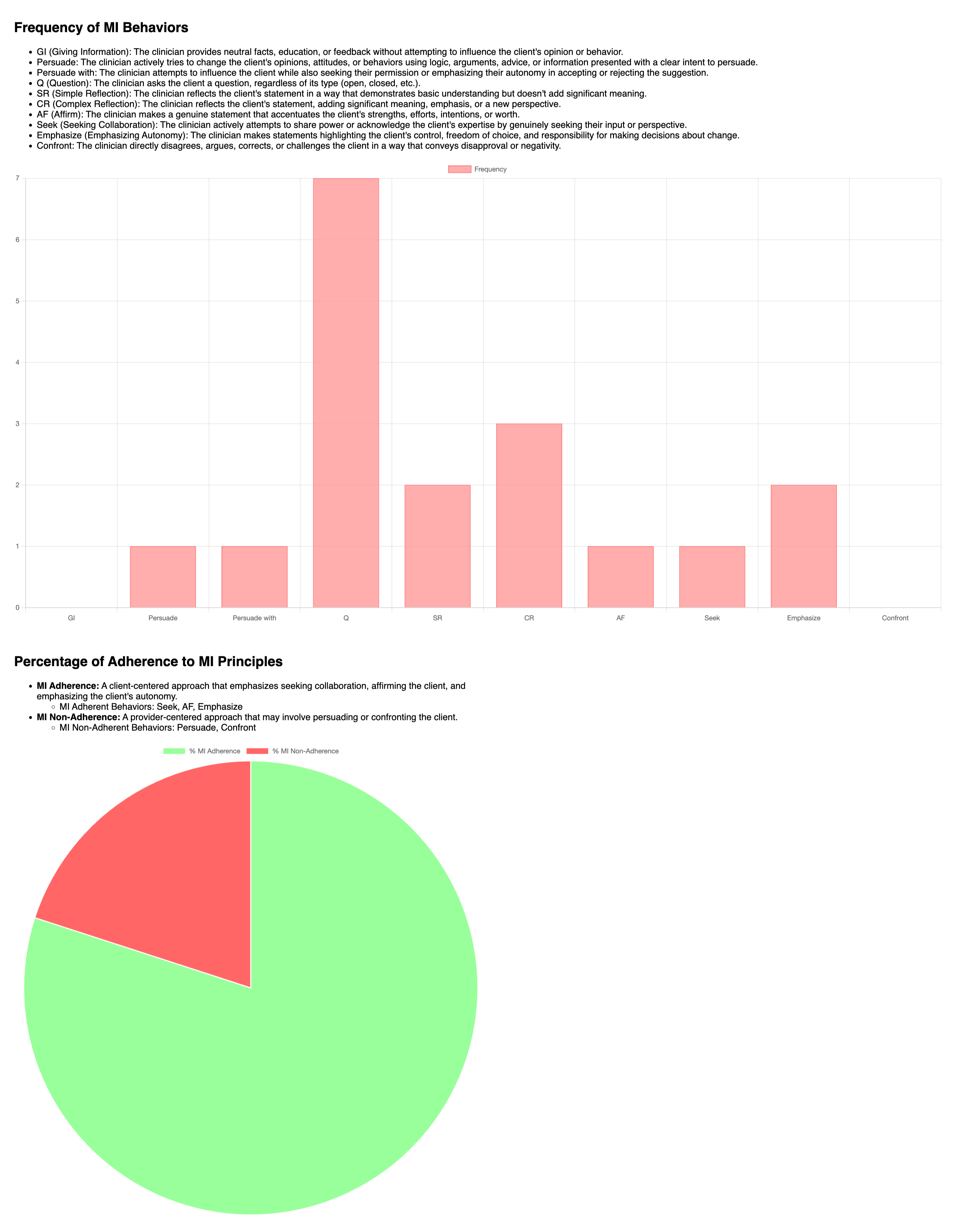}
\end{figure*} 
\begin{figure*}[htb]
\centering
\includegraphics[width=\textwidth]{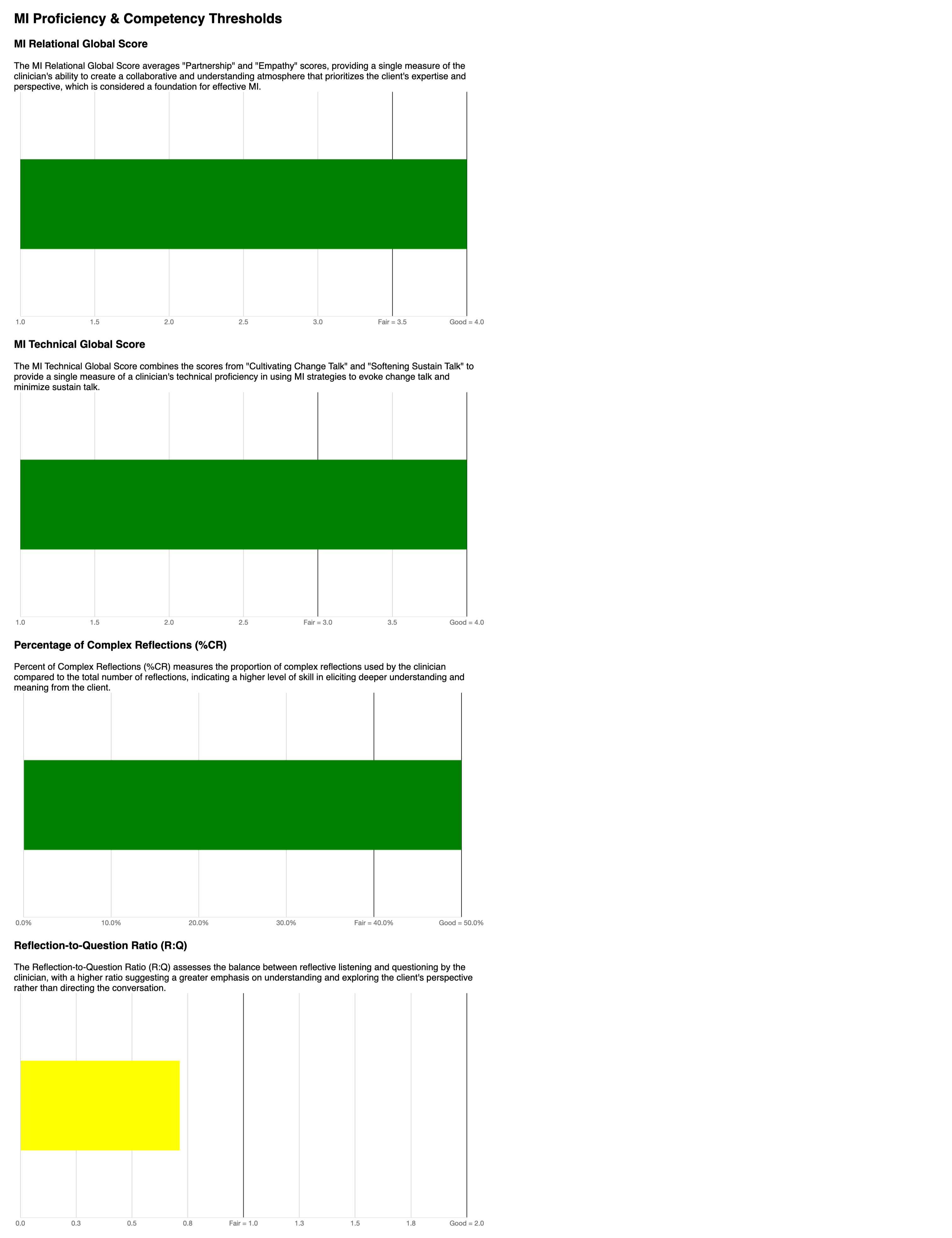}
\end{figure*} 
\begin{figure*}[htb]
\centering
\includegraphics[width=\textwidth]{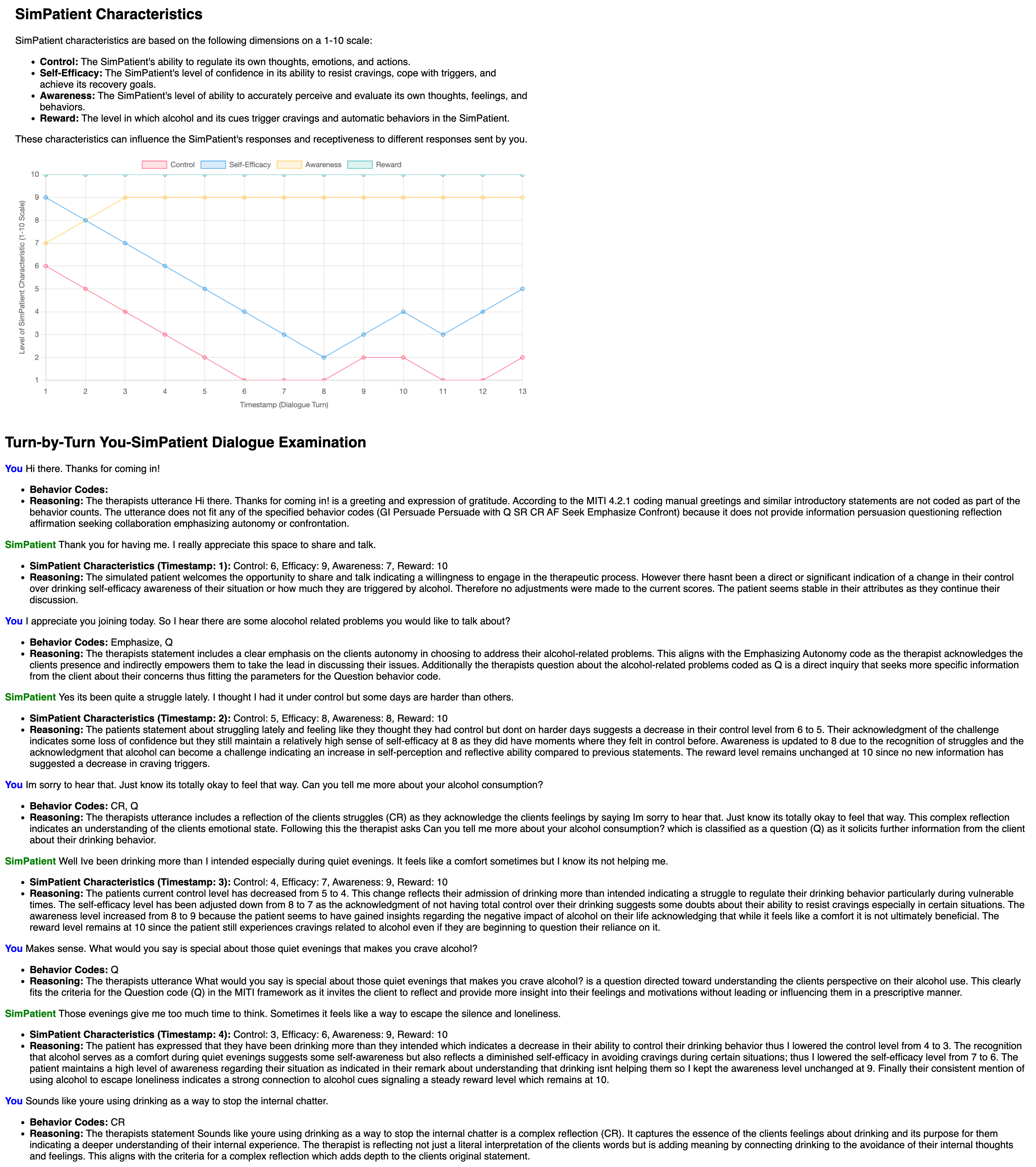}
\end{figure*} 

\end{document}